\documentclass[aps, pra, reprint, floatfix]{revtex4-1}

\usepackage{amsmath}
\usepackage{amssymb}

\usepackage{amsthm}
  \newtheoremstyle{break}
  {}
  {}
  {\itshape}
  {}
  {\bfseries}
  {}
  {\newline}
  {}

  \theoremstyle{break}
  \newtheorem{theorem}{Theorem}
  \newtheorem{definition}{Definition}
  \newtheorem{lemma}[theorem]{Lemma}
  
  \newtheorem{proposition}[theorem]{Proposition}
  
\usepackage[]{graphicx}

\usepackage[colorlinks=false, bookmarksnumbered=true, linkcolor=blue]{hyperref}

\begin{document}  

\title{Coexistence on Reflecting Hyperplane in Generalized Probability Theories}
\author{Masatomo Kobayshi}
\email[E-mail: ]{masatomo.koba@gmail.com}
\affiliation{Department of Nuclear Engineering, Kyoto University, Kyoto daigaku-katsura, Nishikyo-ku, Kyoto, Japan 615-8540}
\date{\today}


\begin{abstract}
  The coexistence of effects in a certain class of generalized probability theories is investigated. The effect space corresponding to an even-sided regular polygon state space has a central hyperplane that contains all the nontrivial extremal effects. The existence of such a hyperplane, called a reflecting hyperplane, is tightly related to the point symmetry of the corresponding state space. The effects on such a hyperplane can be regarded as the (generalized) unbiased effects. A necessary and sufficient condition for a pair of unbiased effects in the even-sided regular polygon theories is presented. This result reproduces a low-dimensional analogue of known results of qubit effects in a certain limit.
\end{abstract}

  \maketitle


\section{Introduction}
  The existence of a pair of observables which are not jointly measurable is one of the most important properties of the quantum theory. It is, however, known that this peculiar property is not specific to the quantum theory in a general framework called Generalized Probability Theories (GPTs)\cite{GPT0}. This framework focuses on the operational aspect of probability theories and is broad enough to include both the classical and quantum theories. According to a recent result obtained by \cite{CE_Plavala16}, a GPT is the classical theory if and only if every pair of observables is jointly measurable.
  
  For a given GPT, it is interesting to study its closeness to the classical theory (and the quantum theory). For this purpose jointly measurable pairs of observables may play important role. Unfortunately, the problem to find any necessary and sufficient conditions for a pair of observables to be jointly measurable seems to be an intractably hard problem even in the quantum theory. An important partial result was given by Busch \cite{CE_Busch86} in 1986. In the paper, he treated the so-called unbiased binary observables whose POVM element (effect) is traceless (i.e., unbiased effect) on a qubit and obtained a simple necessary and sufficient condition for a pair of such observables to be jointly measurable.
  
  In this paper, we study GPTs which have a certain symmetry. The symmetry allows us to introduce a natural generalization of the unbiased binary observable. As we study only binary observables, the notion of jointly measurable observables coincide with that of coexistent effects. The GPTs we treat have even-sided regular polygon theories (GPTs specified by regular polygon state spaces) as its main examples.
  
  We first study the even-sided regular polygon theories and show that the corresponding effect space has a nice symmetry which allows the existence of a hyperplane so that this plane has all the extreme effects except for trivial ones (zero and unit) and divides the whole effect space into two subsets which are related to each other by reflection symmetry. We call such a hyperplane a {\em reflecting hyperplane}. We generalize the notion of unbiased effects to a GPT with a reflecting hyperplane. We obtain a necessary and sufficient condition for a pair of unbiased effects of even-sided regular polygon theories to be coexistent. Furthermore, we show that an effect space has such a reflecting hyperplane if and only if its corresponding state space has the point symmetry. We examine general systems (other than regular polygons) which have the reflecting hyperplane and show that the volume of the set of all effects coexistent with a nontrivial extreme effect is vanishing.
  
\section{Generalized Probability Theory}
  \label{sec:GPT}
  The GPT provides a general frameworks for description of probabilistic physical experiments. We prepare a measured object and perform a measurement to obtain an outcome. A preparation of a measured object is called a {\em state}. Once a state is fixed, a measurement process assigns a probability to each possible outcome. This probability assigning map whose range is $[0,1]$ is called an {\em effect}. As an experiment is determined by possible classes of preparation and measurements, a GPT is specified by a state space $S$ and an effect space $E$.
  
  Any statistical ensemble of states is also a state in quantum mechanics. This property is also applied to any GPTs. That is, if we have two states $\omega^1, \omega^2$ then $\omega = p\omega^1 + (1-p)\omega^2$ (for $0\leq p\leq 1$) is also a state. Since this state $\omega$ is interpreted as a statistical mixture of $\omega_1$ and $\omega_2$, this implies $e(\omega) = pe(\omega^1) + (1-p)e(\omega^2)$ for all effects $e\in E$. The effect space is also closed under the statistical mixture. In other words, for effects $e_1, e_2$ and $0\leq p\leq 1$, their statistical mixture $e = qe_1 + (1-q)e_2 \, (0<q<1)$ becomes an effect. For simplicity, we shall assume that $S$ (and hence $E$) is embeddable into a finite dimensional vector space on $\mathbb{R}$. In addition as we assume that $S$ (and hence $E$) is compact with respect to a natural induced topology and thus has sufficiently many extremal points. An extremal point in $S$ is called a pure state.
  \begin{definition}[GPT]
    A generalized probability theory (GPT) is a doublet $(S,E)$ satisfying the following properties.
    \begin{itemize}
      \item The state space $S$ is a compact convex set embedded in a finite dimensional vector space.
      \item The effect space $E$ is a (convex) sub space of the dual space of $S$ and satisfies $0\leq e(\omega)\leq1$ for all $\omega\in S$.
    \end{itemize}
  \end{definition}

  An affine function $u$ satisfying $u(\omega)=1$ for all $\omega \in S$ is called a unit. Hereafter we assume each effect space has the unit as its element. We call an effect space $E= \{e|\mbox{affine},\, 0\leq e(\omega)\leq 1 \mbox{ for all }\omega \in S \}$ {\em corresponding} to a state space $S$, which is the maximal set as an effect space for a given state space $S$. In this paper, we study $(S, E)$ in which $E$ is corresponding to $S$ unless otherwise stated. We define the measurement with $u$ as follows:
  \begin{definition}
    An observable (measurement) $M$ is a set of effects with $\sum_{m\in M}m = u$.
  \end{definition}
  
  For $e\in E$, its {\em complement} effect $\bar{e}:=u-e$ is also an element of $E$. We call $o := \bar{u}$ a zero effect. Then it is easy to see $E\subset \{e| o \leq e \leq u\}$ where $\leq$ is natural order, i.e. $e_l\leq e_g$ means $e_l(\omega)\leq e_g(\omega)$ for all $\omega\in S$. For any effect $e$, $\frac{1}{2} e + \frac{1}{2} \bar{e}= \frac{1}{2} e + \frac{1}{2}(u-e) = \frac{1}{2}u$ holds. Thus, the effect $\frac{1}{2}u$ is called a {\em center} of effect space.
  
  The lower set of an effect $e$ is often used in this paper. We denote this set by $\underline{E}(e)$, i.e. $\underline{E}(e) := \{e'\in E|o\leq e'\leq e\}$. The upper one is similarly defined, i.e. $\overline{E}(e) := \{e'\in E|e\leq e'\leq u\}$. The following proposition shows an important property of extremal effects.
  \begin{proposition}
    The complement of an extremal effect $e$ is also extremal effect.
    \begin{proof}
      If $\bar{e}$ is not extremal point then there exists $p \in (0,1), f_1,f_2\in E$ such that $\bar{e} = pf_1+(1-p)f_2$. Its complement is $e=p\bar{f_1}+(1-p)\bar{f_2}$, because $\bar{\bar{e}}=e$. This is contradiction to $e$ is extremal.
    \end{proof}
  \end{proposition}
  
\section{Regular Polygon Theories}\label{ssec:ex1}
  In this section we study the structure of effect spaces corresponding to regular polygon state spaces. In particular we will show that the effect space corresponding to every even-sided regular polygon state space has a reflection symmetry. To clarify this point, we first introduce a special hyperplane in an effect space.
  \begin{definition}
    A {\em central hyperplane} is a hyperplane in an effect space which contains the effect $\frac{1}{2}u$. A {\em reflecting hyperplane} is a central hyperplane which contains all extremal effects except for zero and unit effects.
  \end{definition}
  As we will see below, the effect space corresponding to every even-sided regular polygon state space has a reflecting hyperplane. For simplicity, we often use a coordinate $(x^1,x^2, \ldots, x^d)$ of an effect space so that its reflecting hyperplane is $x^d=\frac{1}{2}$.
  
  \subsection{Example 0: Classical Theory}
    The $n$-level classical theory has a state space as a simplex with $n$ vertexes. One of the simple ways to parametrize a simplex in a vector space is to represent its extremal points as $(0,\cdots,0,1,0,\cdots,0)$ (only $j$th element is 1). The corresponding effect space is restricted by $0\leq e(\omega)\leq1$, hence we know it is $\{e=(e^1,e^2,\cdots,e^n) \,|\, 0\leq e^i\leq1\}$. It has the extremal points $(\frac{1}{2}\pm\frac{1}{2}, \frac{1}{2}\pm\frac{1}{2}, \cdots, \frac{1}{2}\pm\frac{1}{2})$, and it shapes hypercube in $n$-dimensional space. Its unit and zero effects are obviously represented as $(1,1,\cdots)$, $(0,0,\cdots)$ respectively.
    
  \subsection{Example 1: Square Bit Space}
    Now we construct an important theory called {\em square bit} (or gbit) by restricting a classical effect space. This method is introduced in [\cite{GPT1}]. We treat the four-level classical effect space, and restrict it to a subset $E_{\rm sb}$ satisfying $e^1+e^2=e^3+e^4$. The extremal points of effect space are reduced to six points as $e_1=(1,0,0,1), e_2=(1,0,1,0), e_3=(0,1,0,1), e_4=(0,1,1,0)$, zero, and unit effects. This new space shares the zero and unit effects with the original, thus $E_{\rm sb}$ is sub effect space. Two states $\omega^1 = (\omega^1_1,\omega^1_2,\omega^1_3,\omega^1_4), \omega^2 = (\omega^2_1,\omega^2_2,\omega^2_3,\omega^2_4)$ in the four-revel classical state space are equivalent (i.e., cannot be distinguished by effects in $E_{\rm sb}$) if
    \begin{equation}
      \omega^2_1 = \omega^1_1 + t,\,\,\, \omega^2_2 = \omega^1_2 + t,\,\,\, \omega^2_3 = \omega^1_3 - t,\,\,\, \omega^2_4 = \omega^1_4 - t
    \end{equation}
    where $t\in \mathbb{R}$, since $e(\omega^1) = e(\omega^2)$ for all effects $e$ in $E_{\rm sb}$. This equivalence also allows us to reduce the dimension of state space because the state is an equivalent class of preparations. We project the original state space to a surface $\omega_1+\omega_2=\frac{1}{2} (=\omega_3+\omega_4)$ and obtain the extremal points, i.e. $\omega^1=(\frac{3}{4},-\frac{1}{4},\frac{1}{4},\frac{1}{4}),\omega^2=(-\frac{1}{4},\frac{3}{4},\frac{1}{4},\frac{1}{4}),\omega^3=(\frac{1}{4},\frac{1}{4},\frac{3}{4},-\frac{1}{4}),\omega^4=(\frac{1}{4},\frac{1}{4},-\frac{1}{4},\frac{3}{4})$. We show the probability table of the square bit theory at TABLE. \ref{tb:gbit}. It shows that this theory is neither classical nor quantum one. These results tell us that the state space is a square and the effect space forms an octahedron. The (normalized) state space lives in a two-dimensional space and the effect space lives in a three-dimensional space.
    \begin{table}[hbt]
      \begin{center}
        \caption{square bit probabilities}
        \begin{ruledtabular}
          \begin{tabular}{c|cccccc}
            $\omega\backslash e$ & $o$ & $e_1$ & $e_2$ & $e_3$ & $e_4$ & $u$ \\ \hline
            $\omega^1$ & 0 & 1 & 0 & 0 & 1 & 1 \\
            $\omega^2$ & 0 & 1 & 0 & 1 & 0 & 1 \\
            $\omega^3$ & 0 & 0 & 1 & 0 & 1 & 1 \\
            $\omega^4$ & 0 & 0 & 1 & 1 & 0 & 1
          \end{tabular}
          \label{tb:gbit}
        \end{ruledtabular}
      \end{center}
    \end{table}
    
  \subsection{Example 2: Regular Polygon Theories} \label{ssec:ex2}
    In this subsection, we generalize the last example to the general two-dimensional regular polygons. We give an explicit parameterization of the state space to study the structure of its corresponding effect space.
    
    We denote states as column vectors, effects as row vectors, and the dual pair with $e\in E$ and $\omega\in S$ as $\langle e\omega\rangle$.
    \subsubsection{State and Effect spaces}
      We define a rotation matrix $R(\theta)$ on $z={\rm (const)}$ surface as
      \begin{eqnarray}
        R(\theta) = \begin{pmatrix}
        \cos\theta & -\sin\theta & 0 \\ 
        \sin\theta & \cos\theta & 0 \\ 
        0 & 0 & 1
        \end{pmatrix}.
      \end{eqnarray}
      The pure states of a $n$-sided regular polygon are defined as $\{\omega^k|\omega^k = R(\frac{2k\pi}{2n})\omega^0, k\in\mathbb{N} \}$, where $\omega^0 = (1, 0, 1)^T$. Thus the state space is a regular polygon whose vertexes have distance $1$ from the origin on the plane $z=1$. The effect space is a set represented by $\{e\,|\,0 \leq \langle e\omega^k\rangle\leq 1\}$. We set $m^0_k:=\{e \in E| \langle e\omega_k\rangle = 0\}$ and $m^1_k :=\{e \in E| \langle e\omega_k\rangle = 1\}$. If the number of pure states is odd, the extremal effects are intersections of $m^0_k, m^1_{k+\frac{n-1}{2}}, m^1_{k+\frac{n+1}{2}}$ and $m^1_k, m^0_{k+\frac{n-1}{2}}, m^0_{k+\frac{n+1}{2}}$. If the number of pure states is even, the extremal effects are intersections of $m^0_k, m^0_{k+1}, m^1_{k+\frac{n}{2}}, m^1_{k+1+\frac{n}{2}}$. The $m_k$s are also symmetric under the rotation, therefore once one of the intersections is calculated the other extremal points are obtained by rotating it properly. The extremal effects is shown in TABLE. \ref{tb:poly}.
      \begin{table*}[hbt]
        \begin{center}
          \caption{The list of extremal points on $n$-sided regular polygon effect space}
          \begin{ruledtabular}
            \begin{tabular}{cc|cc}
              $n$ & series & k=0 & extremal points \\ \hline
              odd  & zero & none & $(0,0,0)$ \\
              odd  & lower & $\frac{1}{1-\cos(\frac{n-1}{n}\pi)}\left(1, 0, -\cos(\frac{n-1}{n}\pi)\right)$ & $\frac{1}{1-\cos(\frac{n-1}{n}\pi)} \left( \cos(\frac{2k}{n}\pi), \sin(\frac{2k}{n}\pi), -\cos(\frac{n-1}{n}\pi) \right)$ \\
              odd  & upper & $\frac{1}{1-\cos(\frac{n-1}{n}\pi)} \left(-1, 0, 1 \right)$ & $\frac{1}{1-\cos(\frac{n-1}{n}\pi)}\left( -\cos(\frac{2k}{n}\pi), -\sin(\frac{2k}{n}\pi), 1 \right)$ \\ 
              odd  & unit & none & $(0,0,1)$ \\ \hline
              even & zero & none & $(0,0,0)$ \\ 
              even & on hyperplane & $\frac{1}{2}\left(1, \tan(\frac{\pi}{n}), 1 \right)$ & $\frac{1}{2}\left( \cos\frac{2k}{n}\pi + \tan(\frac{\pi}{n})\sin(\frac{2k}{n}\pi), -\sin\frac{2k}{n}\pi + \tan(\frac{\pi}{n})\cos(\frac{2k}{n}\pi), 1 \right)$ \\
              even & unit & none & $(0,0,1)$
            \end{tabular}
            \label{tb:poly}
          \end{ruledtabular}
        \end{center}
      \end{table*}
      
      The zero effect is on the origin and unit one is on $(0,0,1)$. For even-sided polygons, there exists a reflecting hyperplane $z=\frac{1}{2}$. We call this parametrization a {\em normal} parametrization. 
      
      The different parametrization can be obtained by affine transformations of the normal parametrization. An affine map on a $d$-dimensional space is described by a ($d+1$)-dimensional augmented matrix, which is consisted from $d\times d$ matrix, zero column $d$-vector and row $(d+1)$-vector with 1 at the end. If an affine map $A$ has its inverse $A^{-1}$ then the dual pairs preserve the value, i.e. $\langle e\omega\rangle=\langle eA^{-1}A\omega\rangle$. Namely, another parametrization on a state space gives the inverse translation to effect space.
      
      For $n=4$ the state space is a rectangle and the effect space forms an octahedron. This means that the four-sided regular polygon theory is equivalent to the square bit space through an affine mapping which translates unit effect to unit effect. Therefore, these two represent the same system. 
      
      The following proposition shows that the existence of a reflecting hyperplane is related to a symmetry of its corresponding state space.
      \begin{proposition}
      \label{prop:ESandSS}
        An effect space has a reflecting hyperplane if and only if its corresponding state space is point symmetric.
        \begin{proof}
          We write the state as the column vector $\omega=(\mathbf{s},1)^T$ and the effect as the row vector $e=(\mathbf{e},e^d)$, where $\mathbf{s}, \mathbf{e}$ are $(d-1)$-dimensional row vector.
          
          We begin with the necessity. Each nontrivial extremal effect is described as $(\mathbf{e},\frac{1}{2})$ and its complement as $(-\mathbf{e},\frac{1}{2})$, thus a state $s$ satisfies $0\leq-\mathbf{e}\cdot\mathbf{s}+\frac{1}{2}\leq1$. This is equivalent to $0\leq\langle(\mathbf{e},\frac{1}{2})(-\mathbf{s},1)^T\rangle\leq1$. Therefore, if $(\mathbf{s},1)^T$ is a state then $(-\mathbf{s},1)^T$ is also a state.
          
          We then show the sufficiency. We label extremal states as $\omega^i=(\mathbf{s}_i,1)^T$. Let us assume that the state space is point symmetric with respect to the origin $({\bf 0}, 1)$. The effects must satisfy the inequalities $0\leq \pm\mathbf{e}\cdot\mathbf{s}_i+e^d \leq 1$. Thus we obtain 
          \begin{equation}
          \label{eq:er}
            -\frac{1}{2}\leq\mathbf{e}\cdot\mathbf{s}_i\leq\frac{1}{2}.
          \end{equation}
          The extremal effects are intersections of subsets of hyperplanes $\pm\mathbf{e}\cdot\mathbf{s}_i+e^d=0 \,\,{\rm or}\,\, 1$. If chosen sets contain only $\pm\mathbf{e}\cdot\mathbf{s}_i+e^d=0$ hyperplane, its corresponding effect is zero effect. Similarly, if they have only $\pm\mathbf{e}\cdot\mathbf{s}_i+e^d=1$ then the effect is unit one, or else we have $\mathbf{e}\cdot\mathbf{s}_i+e^d=0, \, \mathbf{e}\cdot\mathbf{s}_j+e^d=1$ for a pair of $i,j$, hence $\mathbf{e}\cdot\mathbf{s}_j=1+\mathbf{e}\cdot\mathbf{s}_i$. It means $\mathbf{e}\cdot\mathbf{s}_i=-\frac{1}{2}$ from (\ref{eq:er}). Then, all extremal effects are on the hyperplane $e^d=\frac{1}{2}$.
        \end{proof}
      \end{proposition}
      
    \subsubsection{Quantum limit}\label{sssec:QL}
      A quantum bit (qubit) state and its effect spaces are both represented in Bloch spheres. That is, these spaces are $\mathcal{S}(\mathbb{C}^2) = \{\rho\in\mathcal{T}(\mathbb{C}^2) \,\, | \,\, \rho=\frac{1}{2}(I+\vec{r_s}\cdot\vec{\sigma}), \,\, \|\vec{r_s}\|\leq1 \}$ and $\mathcal{E}(\mathbb{C}^2) = \{\rho\in\mathcal{T}_s(\mathbb{C}^2) \,\, | \,\, e=\frac{1}{2}(\alpha I+\vec{r_e}\cdot\vec{\sigma}), \,\, \|\vec{r_e}\|\leq1, \,\, \|\vec{r_e}\|\leq\alpha\leq2-\|\vec{r_e}\|\}$ where $\vec{\sigma}=(\sigma_x, \sigma_y, \sigma_z)$ is a vector composes Pauli matrices. To see the connection between this qubit system and the regular polygon theories, we neglect $z$-direction. That is, we consider a quantum-like system whose $\vec{r}$ is only on the span of $\sigma_x, \sigma_y$. In this situation, the state and effect spaces are regarded as the circle and the bi-cone respectively. Note the effect space has parameter $\alpha$ to radius direction, then the effect spaces vertexes are at $(x,y,\alpha)=(0,0,0), (0,0,2)$.
      
      On the other hand, the state spaces of regular polygon theories are close to a circle whose radius is $1$ at the limit of $n\rightarrow\infty$. Similarly, the effect space close to a bi-cone whose vertex on $(0,0,0), (0,0,1)$. It, however, does not coincide with former one while is homomorphic. This difference occurs from how we earn the probability from state and effect. The probability is calculated as $\frac{1}{2}(\alpha+\vec{r_s}\cdot\vec{r_e})$ in quantum theory with the Bloch sphere, and it is just $\langle e\omega \rangle$ in normally parametrized GPTs. For reducing this differences, we translate the spaces so that $\omega$ equals the state vector $\vec{r_s}$ with the added element 1 at the end, and $e$ equals the half of the effect vector $\vec{r_e}$ with $\alpha$ at the end. Finally, through these reduction and translation, the limit of $n\rightarrow\infty$ is regarded as quantum limit.

\section{Coexistence}
\label{sec:CE}
  Two quantum effects are called {\em coexistent} if one can obtain them in a single measurement scheme. While the study on coexistent effects has a long history \cite{CE_review07}, their complete characterization has not been known yet. On a qubit system, after a beautiful result on unbiased effects by Busch \cite{CE_Busch86}, the complete characterization for general effects have been obtained \cite{CE_Stano08,CE_Busch10}. We study a possible generalization of the former result in GPTs.
  
  \subsection{Definition and Known Theorem}
    The notion of coexistence can be easily translated to GTPs.
    \begin{definition}
      Effects $e,f$ are {\em coexistent} if and only if there exists an observable $G$ and its sub sets $G_e,G_f$ such that
      \begin{equation}
        e = \sum_{e'\in G_e}e', \,\, f = \sum_{e'\in G_f}e'.
      \end{equation}
    \end{definition}
    If $G$ is coexistent observable for $e,f$ then we denote
    \begin{eqnarray}
      g_1 &=& \sum_{e\in G_e\cap G_f}e, \\
      g_2 &=& \sum_{e\in G_e\cap(G\setminus G_f)}e, \\
      g_3 &=& \sum_{e\in (G\setminus G_e)\cap G_f}e, \\
      g_4 &=& \sum_{e\in (G\setminus G_e)\cap(G\setminus G_f)}e.
    \end{eqnarray}
    Thus, one can show that effects $e$ and $f$ are coexistent if and only if there exist effects $g_1, g_2, g_3$ satisfying
    \begin{eqnarray}
      g_1+g_2 &=& e, \label{eq:ce1}\\
      g_1+g_3 &=& f, \label{eq:ce2}\\
      (o\preceq)g_1+g_2+g_3 &\preceq& u. \label{eq:ce3}
    \end{eqnarray}
    
    For a pair of unbiased effects, Busch obtained a simple characterization.
    \begin{theorem}
    \label{th:qce}
      Let us consider a qubit system. For $\vec{\lambda}$ with $\| \vec{\lambda}\| \leq 1$, we define an unbiased effect $e(\vec{\lambda}) := \frac{1}{2}(I+\vec{\lambda}\cdot\vec{\sigma})$. Two unbiased effects $e(\vec{\lambda_1}), e(\vec{\lambda_2})$ are coexistent if and only if \begin{equation}
        \frac{1}{2}\|\vec{\lambda_1}+\vec{\lambda_2}\|+\frac{1}{2}\|\vec{\lambda_1}-\vec{\lambda_2}\| \leq 1. 
      \end{equation}
    \end{theorem}
    If we fix $\vec{\lambda_1}$ then the region which $\vec{\lambda_2}$ is coexistent with $\vec{\lambda_1}$ forms a spheroid. Its singular focuses are $\vec{\lambda_1}$ and $ -\vec{\lambda_1}$. Through the translation from quantum theory to the normal parametrization, shown in \ref{sssec:QL}, the criteria is equivalent to an inequality,
    \begin{equation}
      \|(e_x,e_y)+(f_x,f_y)\|+\|(e_x,e_y)-(f_x,f_y)\| \leq 1, \label{eq:qce_mod}
    \end{equation}
    where $e, f$ are corresponding effects to $e(\vec{\lambda_1}), e(\vec{\lambda_2})$ respectively. Therefore, we know the region with fixed effect $e$ is an ellipse whose singular focuses are at $e, -e$, and it is in a circle whose radius is $\frac{1}{2}$ on $z=\frac{1}{2}$.
    
  \subsection{Necessary and Sufficient Condition for Unbiased Effects to be Coexistent in Regular Polygon Theories}
    In this section we study even-sided regular polygon. The coexistence problem for a proper class of effects becomes treatable due to the existence of the reflecting hyperplane. An effect is called {\em unbiased} if it is on the reflecting hyperplane.
    
    At first, we need a mathematical preparation. We often calculate the regions of regular polygons whose edges have distance $l$ from their origins located at $\mathbf{x}_0$. We consider the region divided by a line located $l$ away from point $\mathbf{x}_0$ at an angle $\theta$ with the line $x=0$. Here, we assume $\mathbf{x}_0 = \mathbf{0}$. If a vector $\mathbf{x}$ is in the region, the inner product with $\mathbf{e}_\theta = (\cos\theta, \sin\theta)$ is less than $l$, i.e. $\mathbf{e}_\theta\cdot\mathbf{x} \leq \lambda$. Shifting $\mathbf{x}_0$ from the origin, we acquire following inequality $\mathbf{e}_\theta\cdot(\mathbf{x}-\mathbf{x}_0) \leq l$. The polygon sector we need is intersection of the regions where $\theta=\frac{2k}{n}\pi$, therefore we obtain a set, 
    \begin{equation}
      R_l(\mathbf{x_0}):=\{\mathbf{x}\,\, | \,\, \mathbf{e}_\frac{2k}{n}\pi\cdot(\mathbf{x}-\mathbf{x}_0) \leq l, \,\, k\in\mathbb{N}\}.
    \end{equation}
    
    Now, we show the coexistent criteria on the reflecting hyperplane in even-sided regular polygon theories.
    \begin{theorem}[coexistent criteria]
    \label{th:ce}
      Let $n$ be an even number. For an $n$-sided regular polygon theory, two unbiased effects $e,f$ are coexistent if and only if they satisfy the inequality for all $k_+, k_- \in \mathbb{N}$,
      \begin{widetext}
        \begin{equation}
          \left\{-\sin\left(\frac{k_+}{n}\pi\right)e_x+\cos\left(\frac{k_+}{n}\pi\right)e_y\right\}\sin\left(\frac{k_-}{n}\pi\right) + \left\{-\cos\left(\frac{k_+}{n}\pi\right)f_x+\sin\left(\frac{k_+}{n}\pi\right)f_y\right\}\cos\left(\frac{k_-}{n}\pi\right) \leq \frac{1}{2}, \label{eq:criteria1}
        \end{equation}
      \end{widetext}
      where $e_x, e_y, f_x, f_y$s are coordinates of $e,f$ under the normal parametrization.
      \begin{proof}
        If $e$ and $f$ are coexistent, then the relations (\ref{eq:ce1},\ref{eq:ce2},\ref{eq:ce3}) hold. They are equivalent to the existence of $g_1, g_3$ satisfying $g_1\in\underline{E}(e), g_3\in\underline{E}(\bar{e})$. Therefore, we need to calculate the Minkowski's sum of $\underline{E}(e), \underline{E}(\bar{e})$ to obtain the coexistent region with $f$. $f$ is on the plane $z=\frac{1}{2}$, since we use the normal parametrization. Thus, the third parameters of $g_1, g_3$ satisfy $g_1^z+g_3^z=\frac{1}{2}$. The cross section of $\underline{E}(e)$ and plane $g_1^z=l$ is intersection of $R_{\frac{1}{2}l}(e_x, e_y)$ and $R_{\frac{1}{2}(1-l)}(0,0)$. Similarly, the cross section of $\underline{E}(\bar{e})$ and plane $g_3^z=\frac{1}{2}-l$ is intersection of $R_{\frac{1}{2}l}(0,0)$ and $R_{\frac{1}{2}(1-l)}(-e_x, -e_y)$. (See FIG. \ref{fig:proof})
        \begin{figure}[tbhp]
          \centering
          \includegraphics[clip, width=8.5cm]{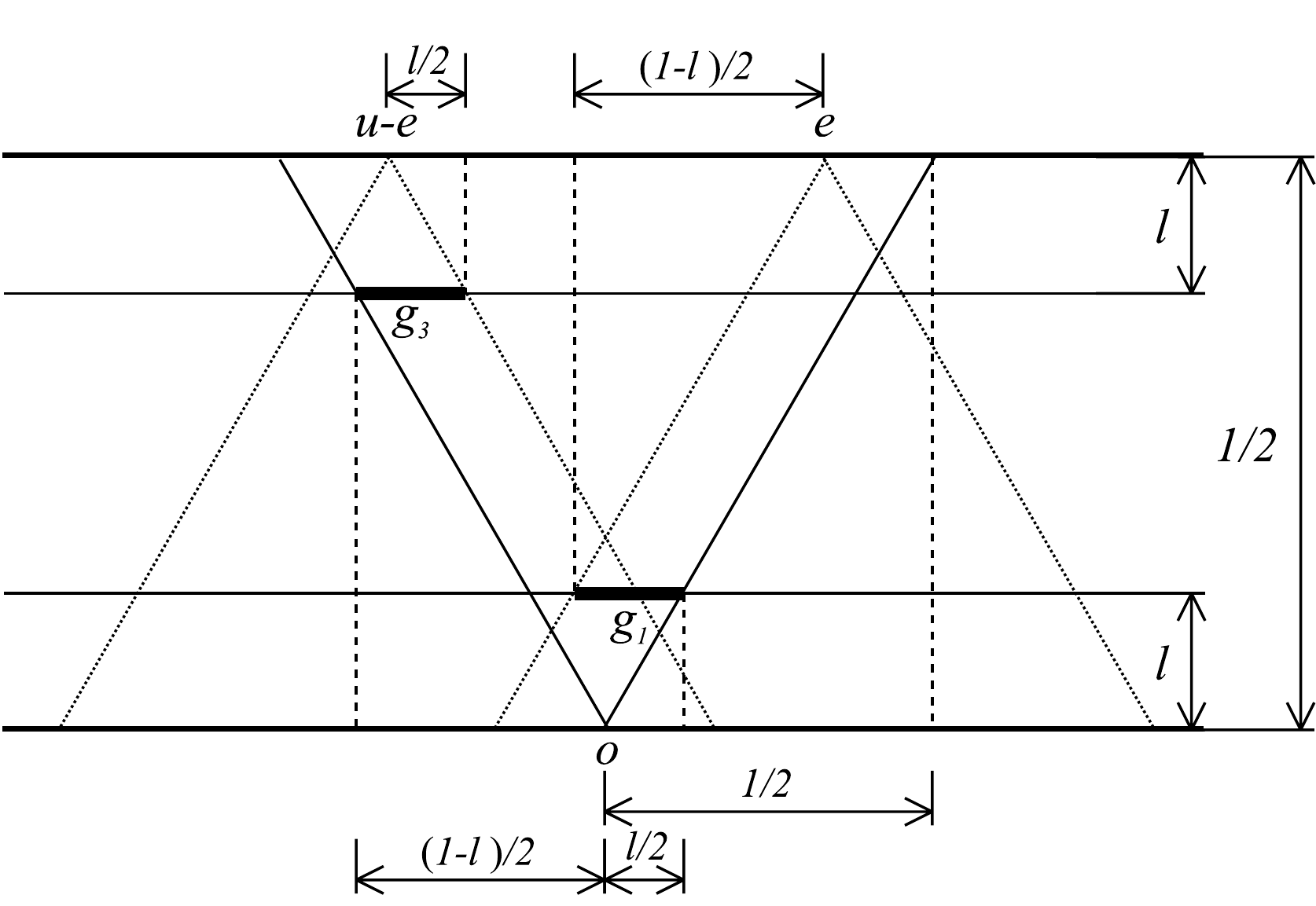}
          \caption{The region where $g_1$ and $g_3$ can be exist on $z=l$ or $z=\frac{1}{2}-l$ planes.}
          \label{fig:proof}
        \end{figure}
        Since these two regions are equivalent under a parallel translation, their Minkowski's sum coincides with the intersection of $R_{l}(\frac{1}{2}e_x,\frac{1}{2}e_y)$ and $R_{1-l}(-\frac{1}{2}e_x,-\frac{1}{2}e_y)$. Thus, the following inequalities hold, 
        \begin{eqnarray}
          (f_x-e_x)\cos\frac{2k}{n}\pi+(f_y-e_y)\sin\frac{2k}{n}\pi &\leq& l  \label{eq:ce_con1}\\
          (f_x+e_x)\cos\frac{2k}{n}\pi+(f_y+e_y)\sin\frac{2k}{n}\pi &\leq& 1-l.  \label{eq:ce_con2}
        \end{eqnarray}
        We obtain the similar inequalities for a fixed $f$. They satisfy an inequality below, and conversely if the following is satisfied then $l$ such that it satisfies (\ref{eq:ce_con1},\ref{eq:ce_con2}) exists.
        \begin{eqnarray}
          (f_x-e_x)\cos\frac{2k_1}{n}\pi+(f_y-e_y)\sin\frac{2k_1}{n}\pi &\leq \nonumber\\ 1-(f_x+e_x)\cos\frac{2k_2}{n}\pi-(f_y+e_y)\sin\frac{2k_2}{n}\pi& \label{eq:ce_eq}
        \end{eqnarray}
        This inequality is equivalent to (\ref{eq:criteria1}), where $k_+=k_1+k_2, k_-=k_1-k_2$.
      \end{proof}
    \end{theorem}
    
\begin{figure*}[tbhp]
      \begin{center}
        \begin{tabular}{c}
          \begin{minipage}{0.243\hsize}
            \begin{center}
              \includegraphics[clip, width=4.3cm]{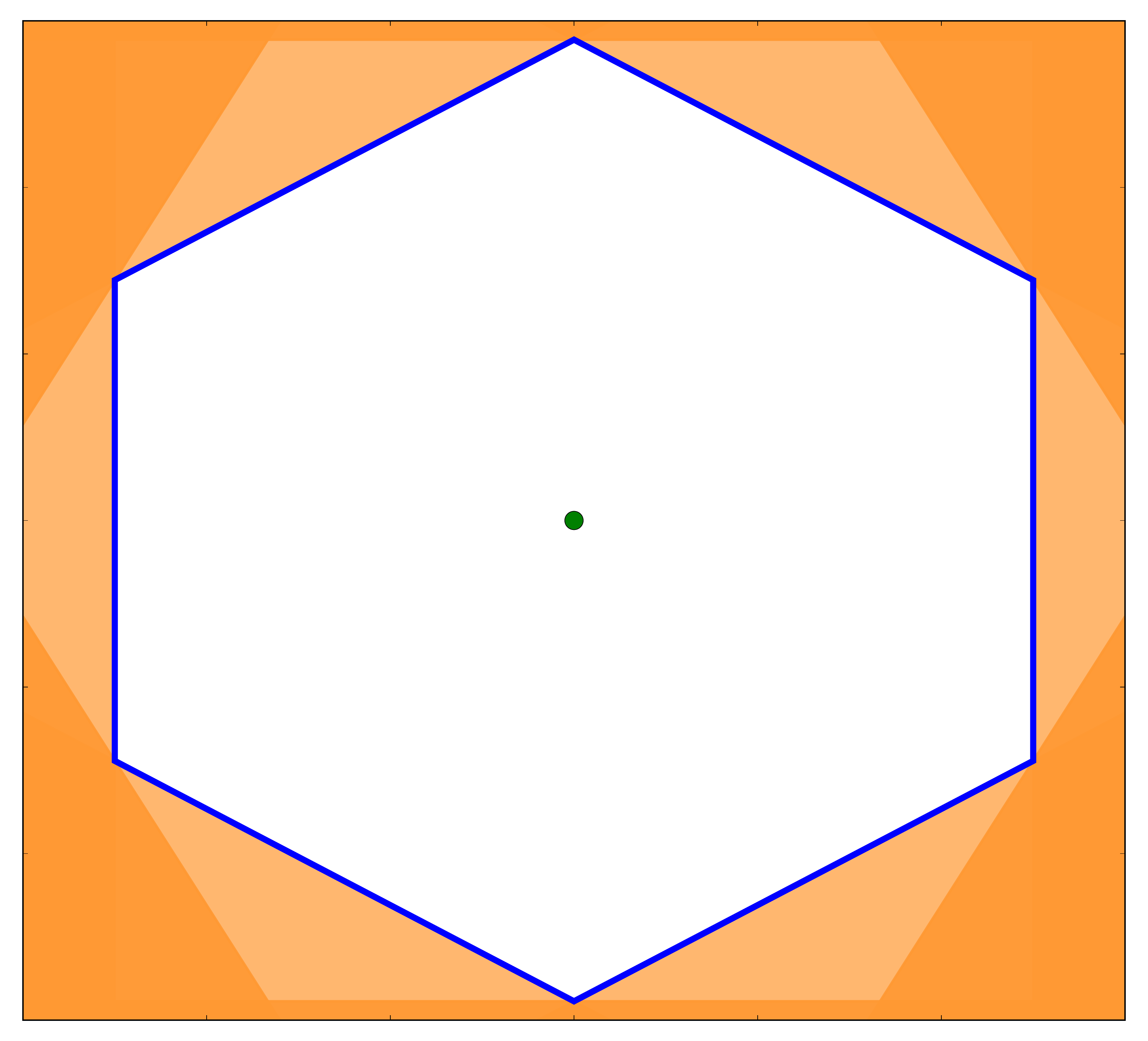}
              \hspace{1.8cm} 2a. [6, 0, 0]
            \end{center}
          \end{minipage}
          \begin{minipage}{0.243\hsize}
            \begin{center}
              \includegraphics[clip, width=4.3cm]{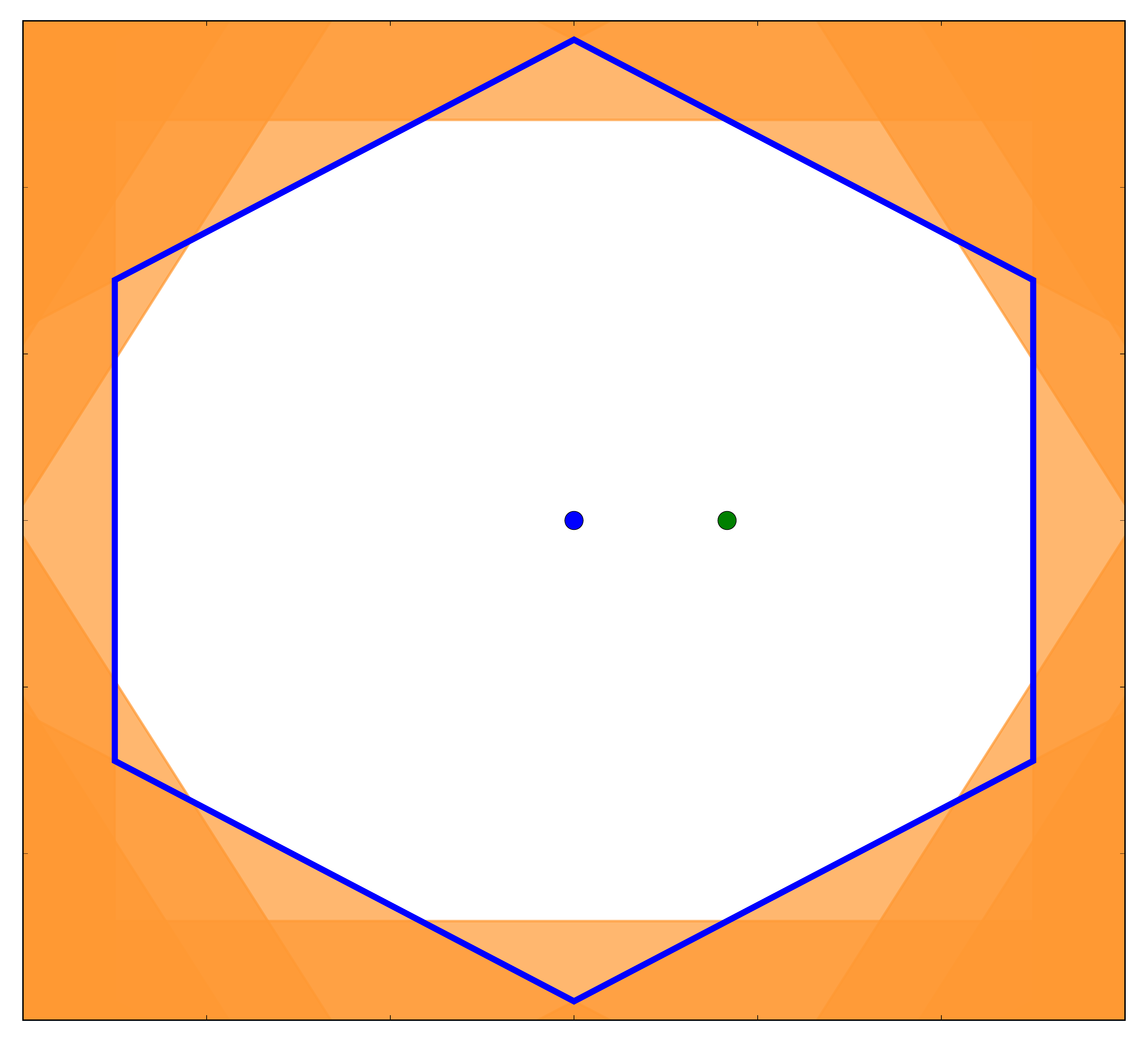}
              \hspace{1.8cm} 2b. [6, 1/3, 0]
            \end{center}
          \end{minipage}
          \begin{minipage}{0.243\hsize}
            \begin{center}
              \includegraphics[clip, width=4.3cm]{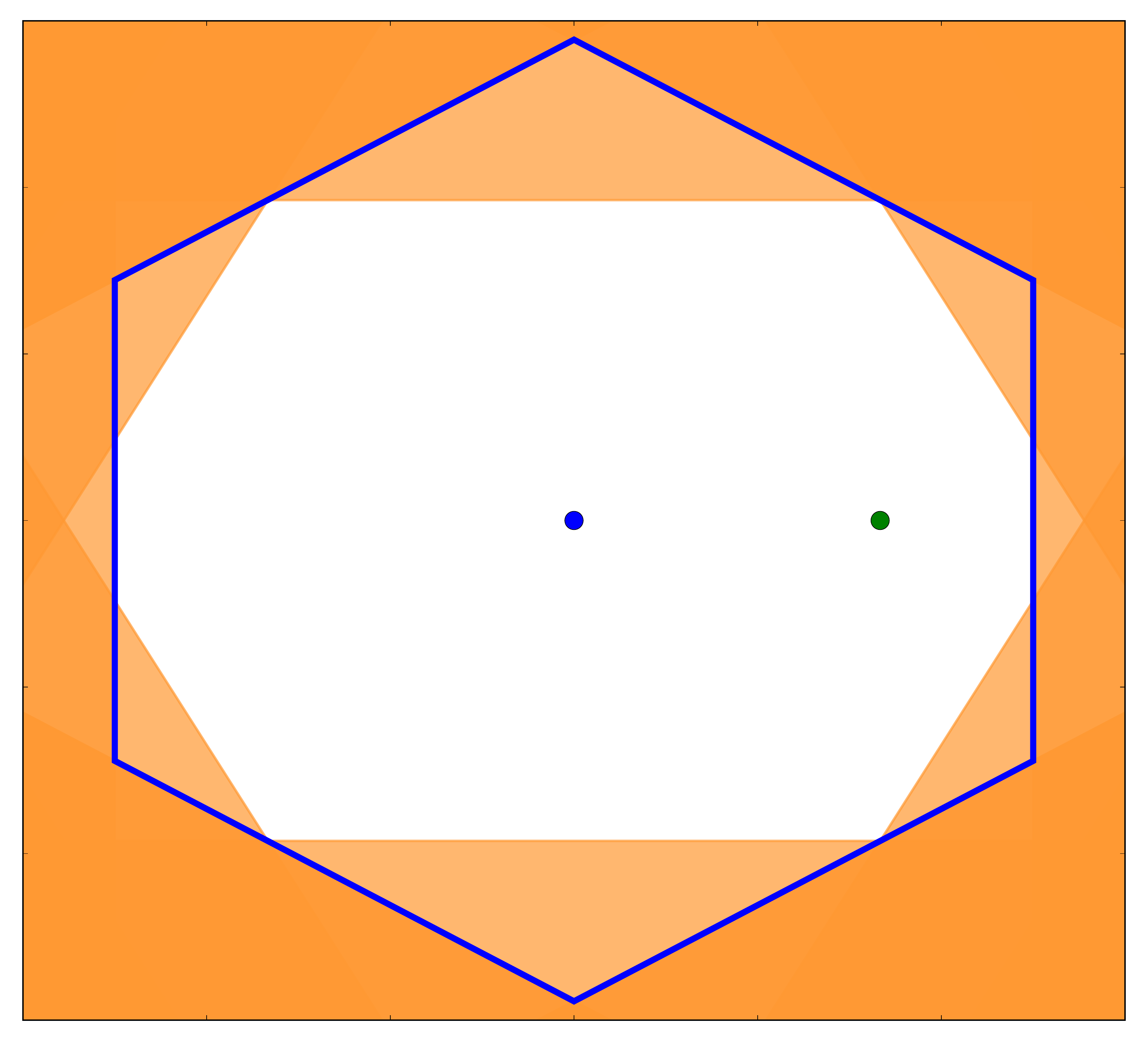}
              \hspace{1.8cm} 2c. [6, 2/3, 0]
            \end{center}
          \end{minipage}
          \begin{minipage}{0.243\hsize}
            \begin{center}
              \includegraphics[clip, width=4.3cm]{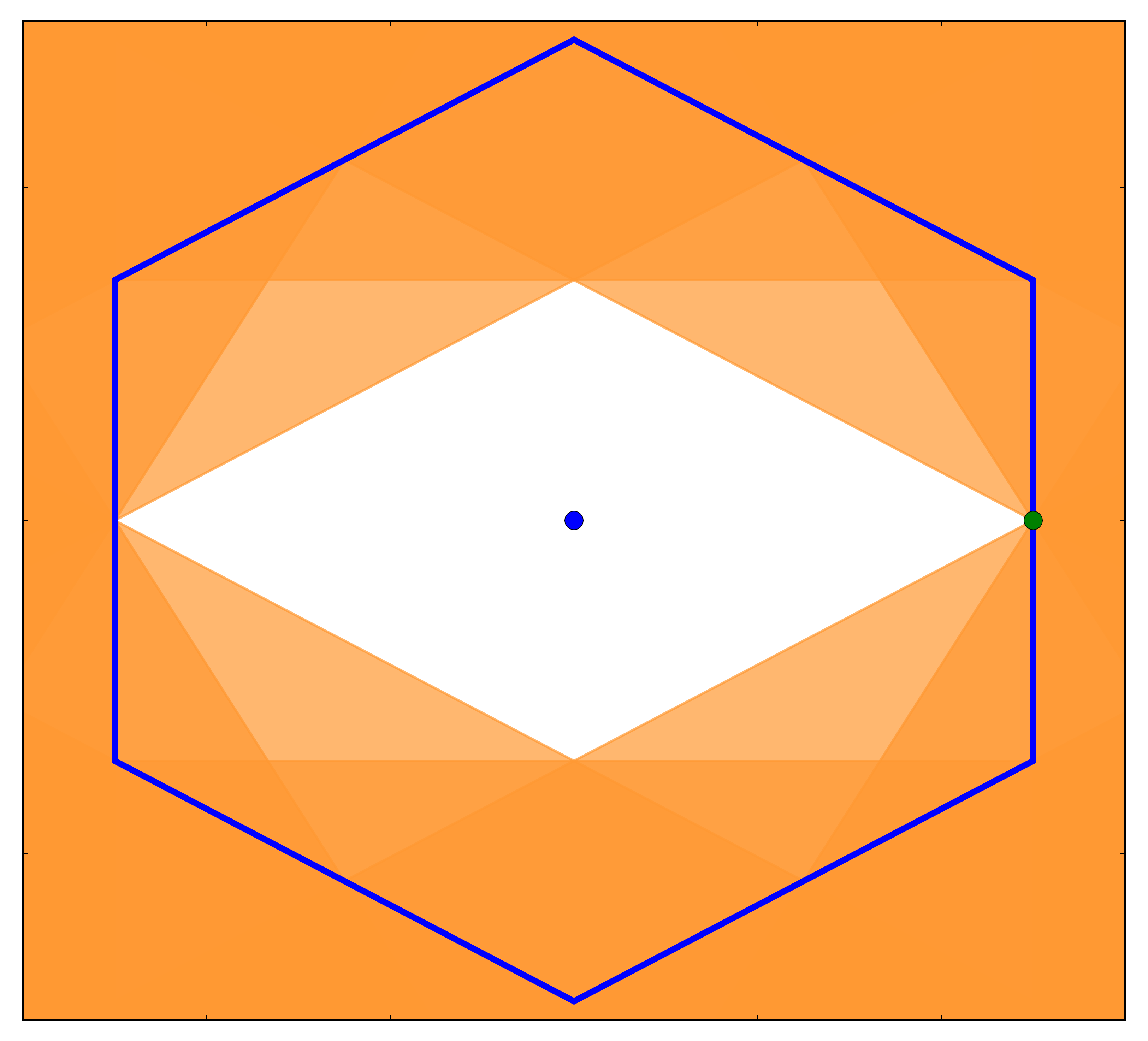}
              \hspace{1.8cm} 2d. [6, 1, 0]
            \end{center}
          \end{minipage} \\
          \begin{minipage}{0.243\hsize}
            \begin{center}
            \end{center}
          \end{minipage}
          \begin{minipage}{0.243\hsize}
            \begin{center}
              \includegraphics[clip, width=4.3cm]{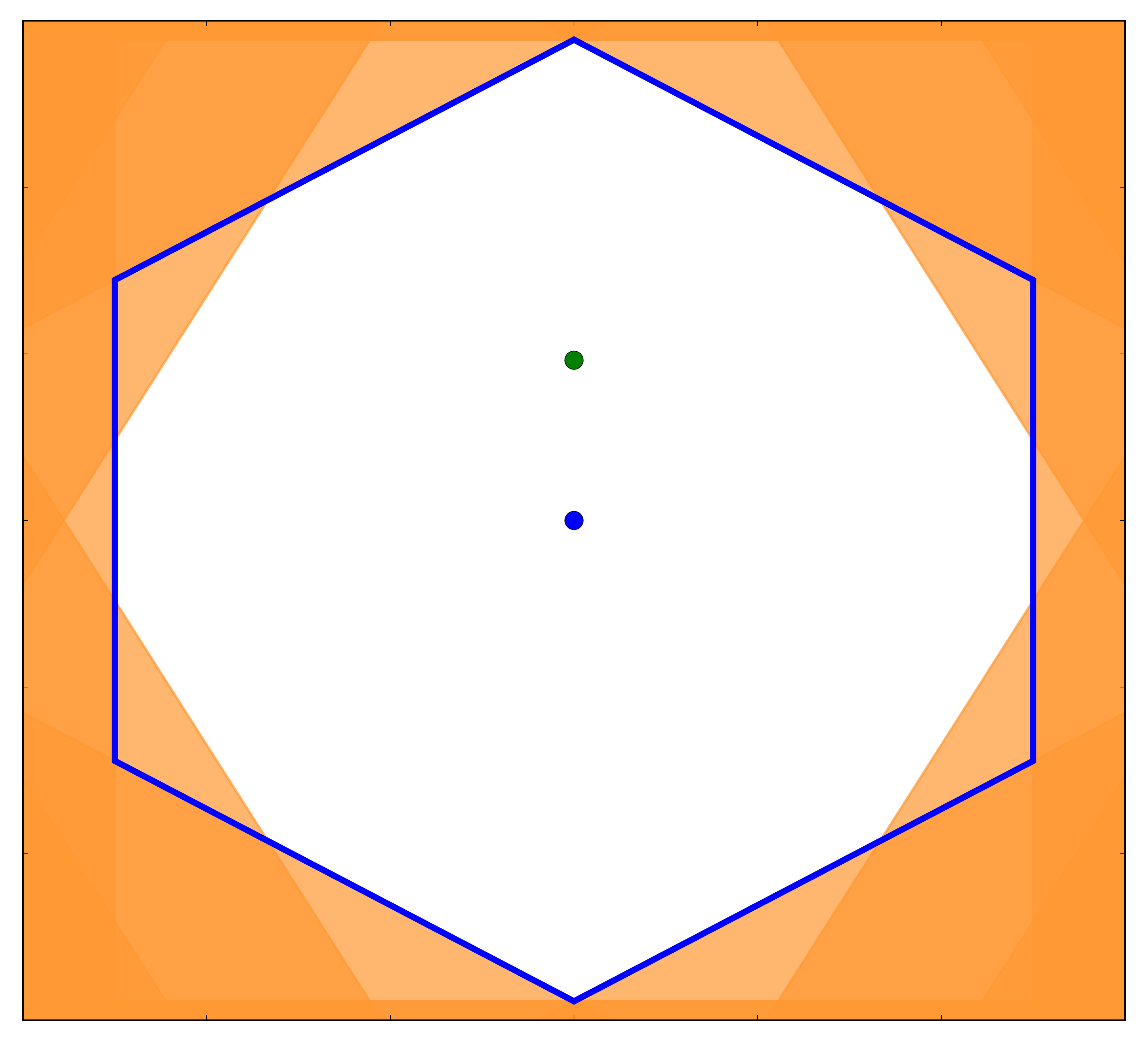}
              \hspace{1.8cm} 2e. [6, 0, 1/3]
            \end{center}
          \end{minipage}
          \begin{minipage}{0.243\hsize}
            \begin{center}
              \includegraphics[clip, width=4.3cm]{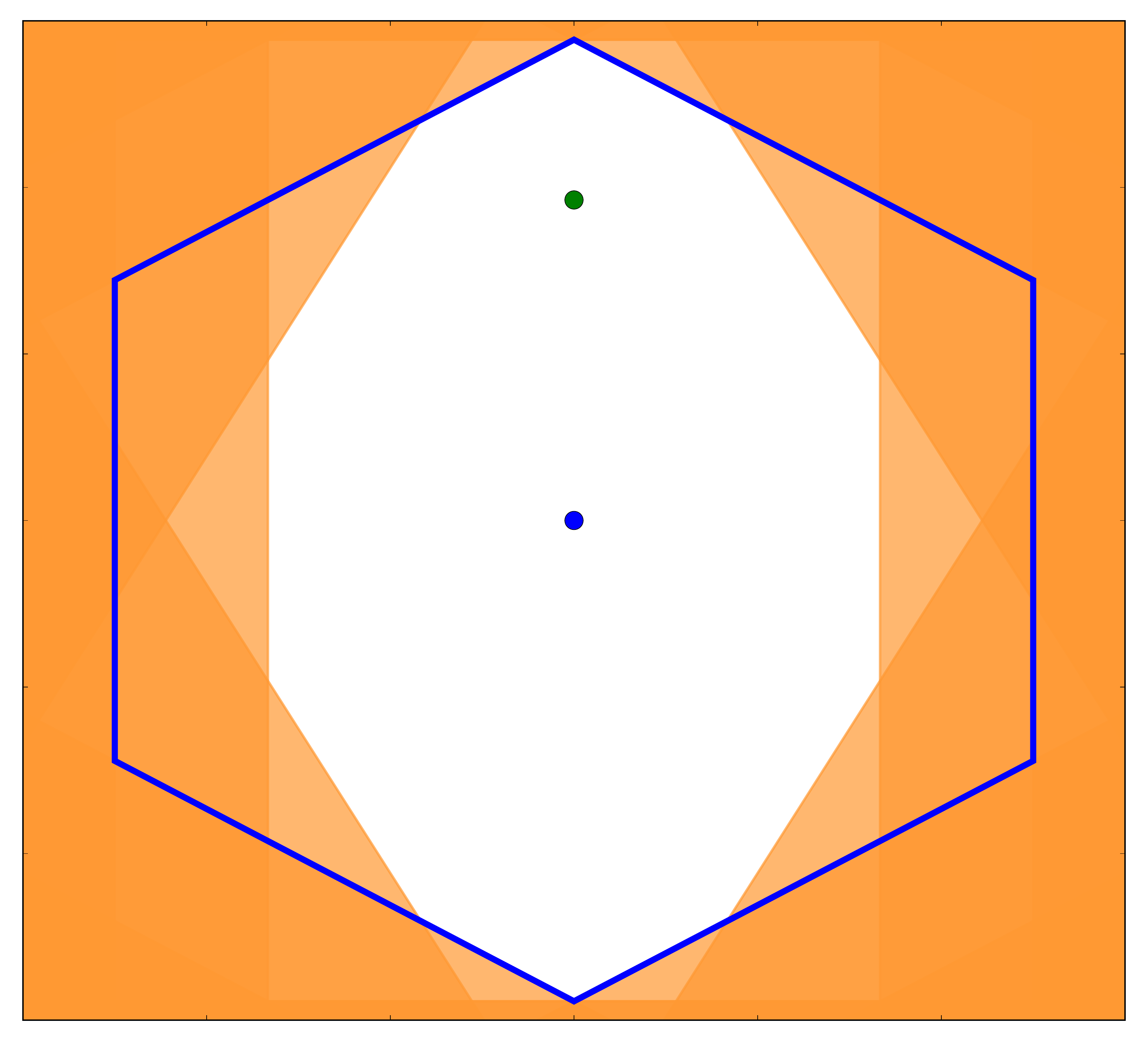}
              \hspace{1.8cm} 2f. [6, 0, 2/3]
            \end{center}
          \end{minipage}
          \begin{minipage}{0.243\hsize}
            \begin{center}
              \includegraphics[clip, width=4.3cm]{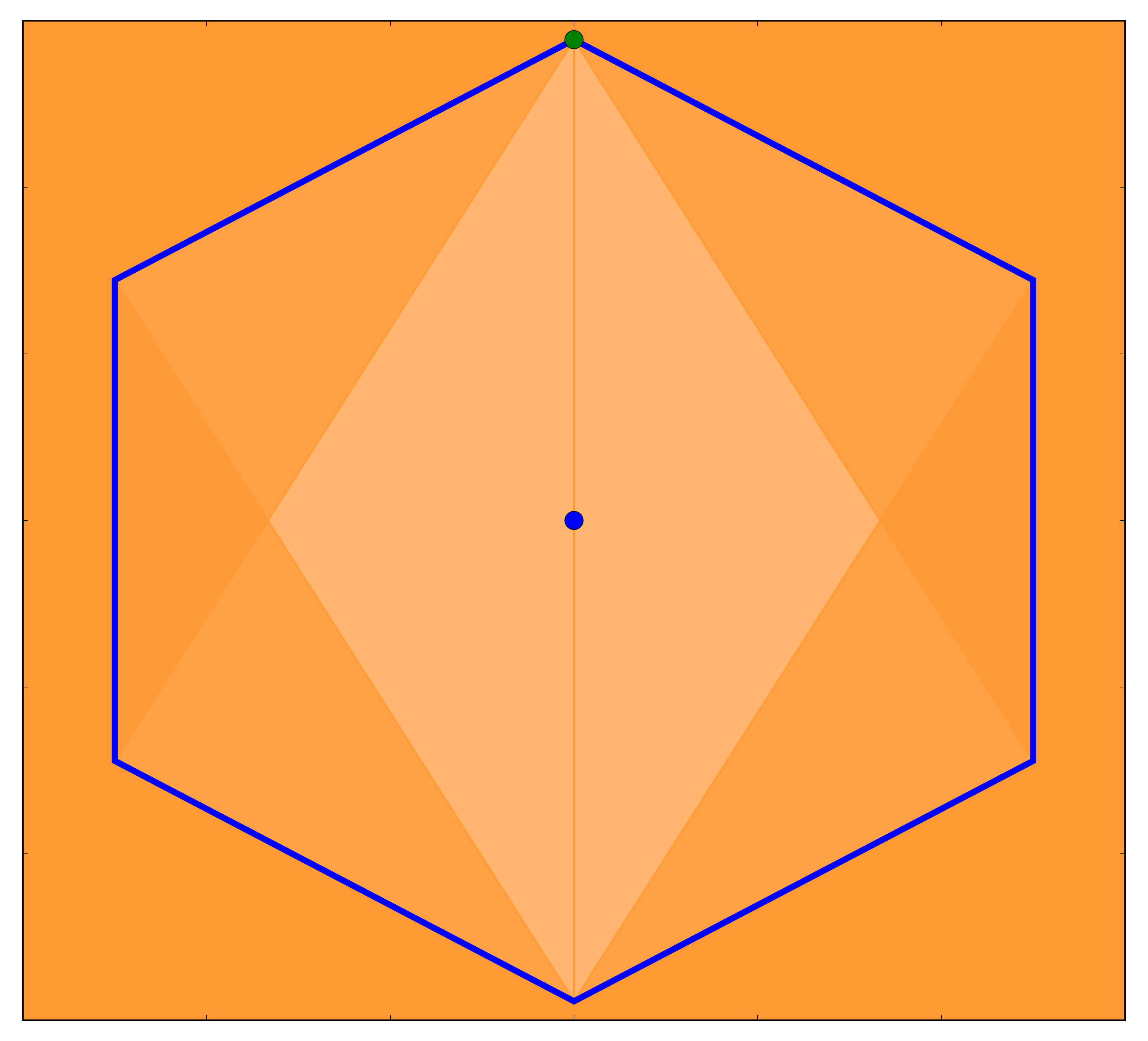}
              \hspace{1.8cm} 2g. [6, 0, 1]
            \end{center}
          \end{minipage}\\
          \begin{minipage}{0.243\hsize}
            \begin{center}
              \includegraphics[clip, width=4.3cm]{figure_620.pdf}
              \hspace{1.8cm} 2h. [6, 2/3, 0]
            \end{center}
          \end{minipage}
          \begin{minipage}{0.243\hsize}
            \begin{center}
              \includegraphics[clip, width=4.3cm]{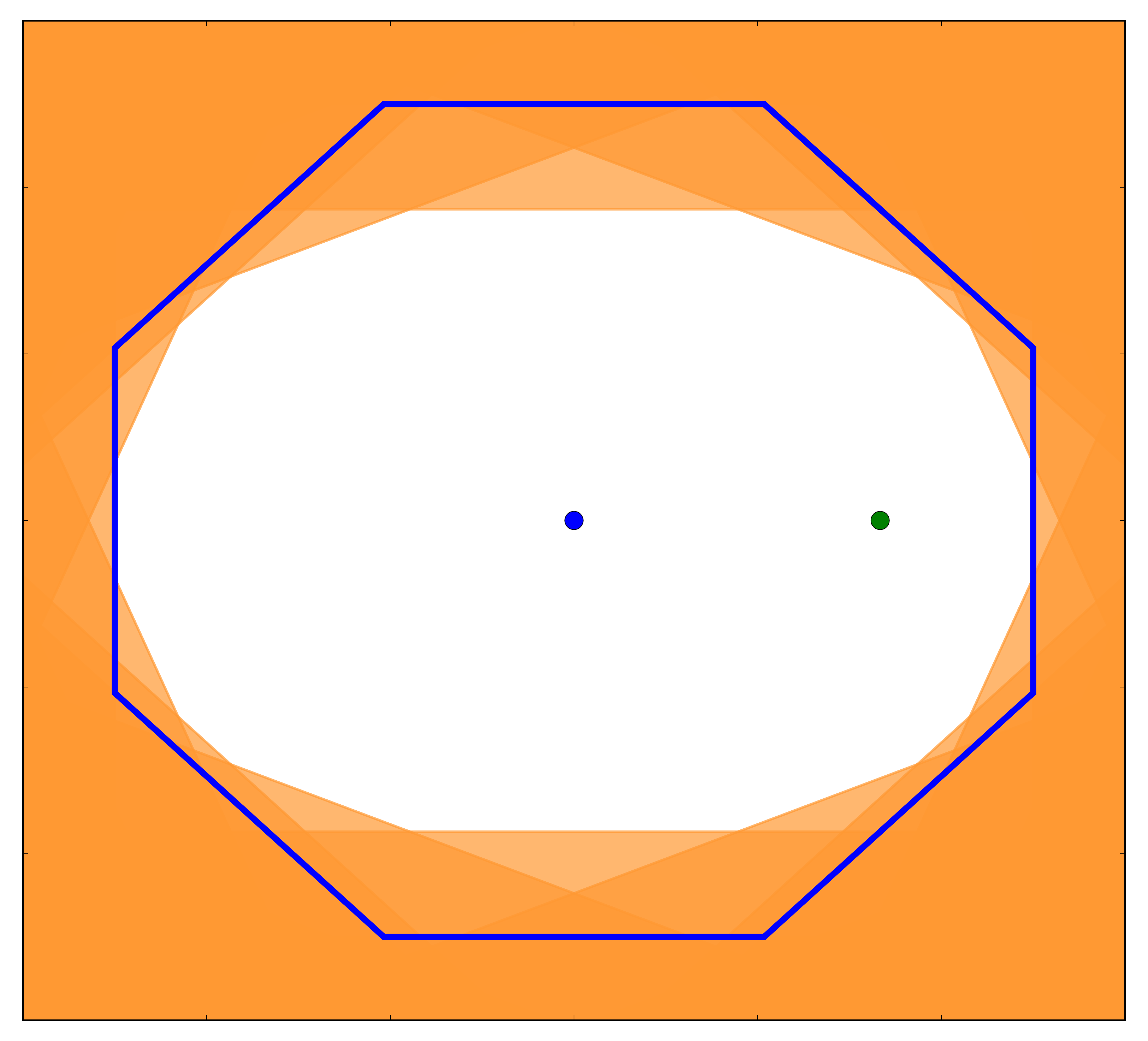}
              \hspace{1.8cm} 2i. [8, 2/3, 0]
            \end{center}
          \end{minipage}
          \begin{minipage}{0.243\hsize}
            \begin{center}
              \includegraphics[clip, width=4.3cm]{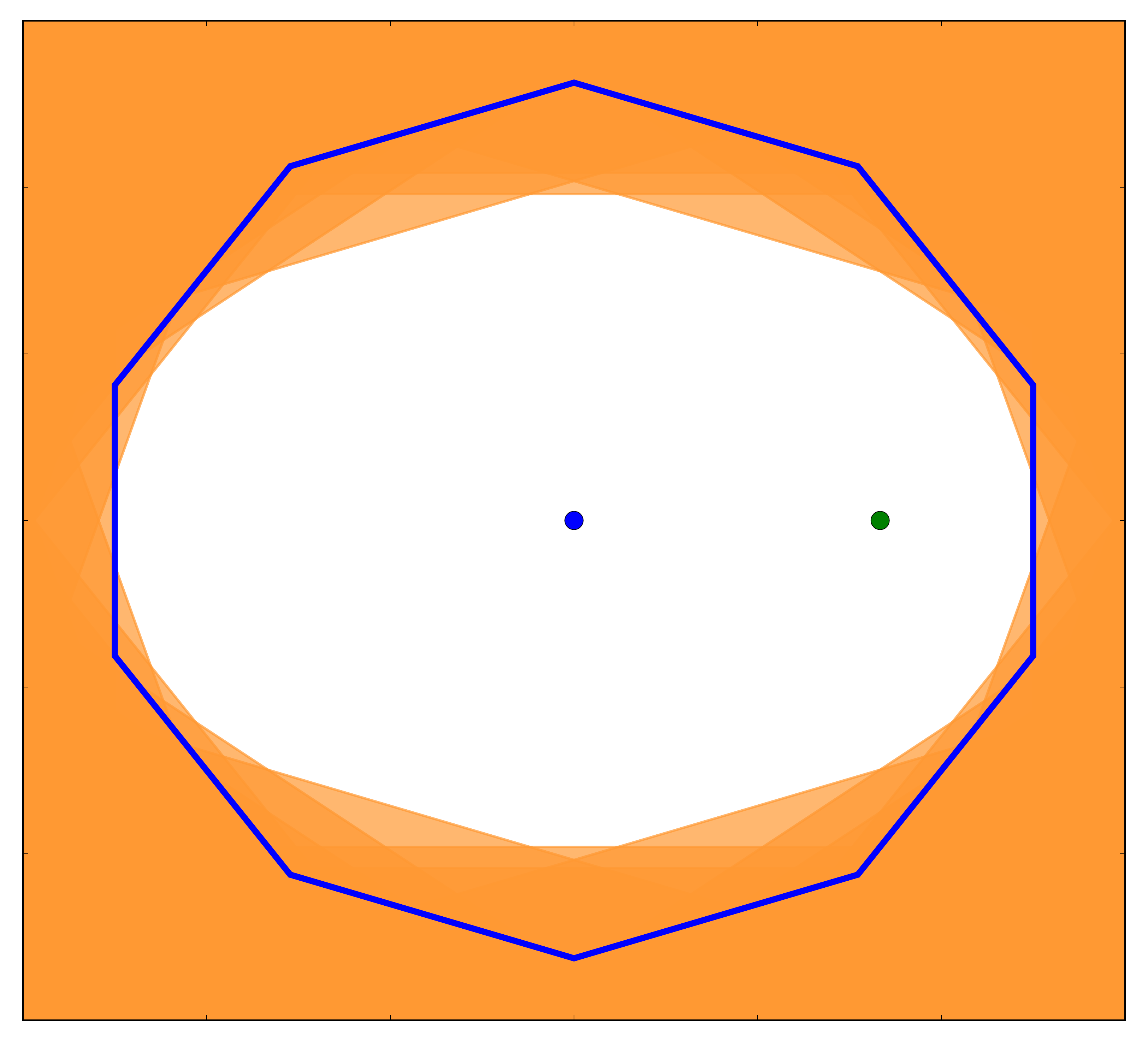}
              \hspace{1.8cm} 2j. [10, 2/3, 0]
            \end{center}
          \end{minipage}
          \begin{minipage}{0.243\hsize}
            \begin{center}
              \includegraphics[clip, width=4.3cm]{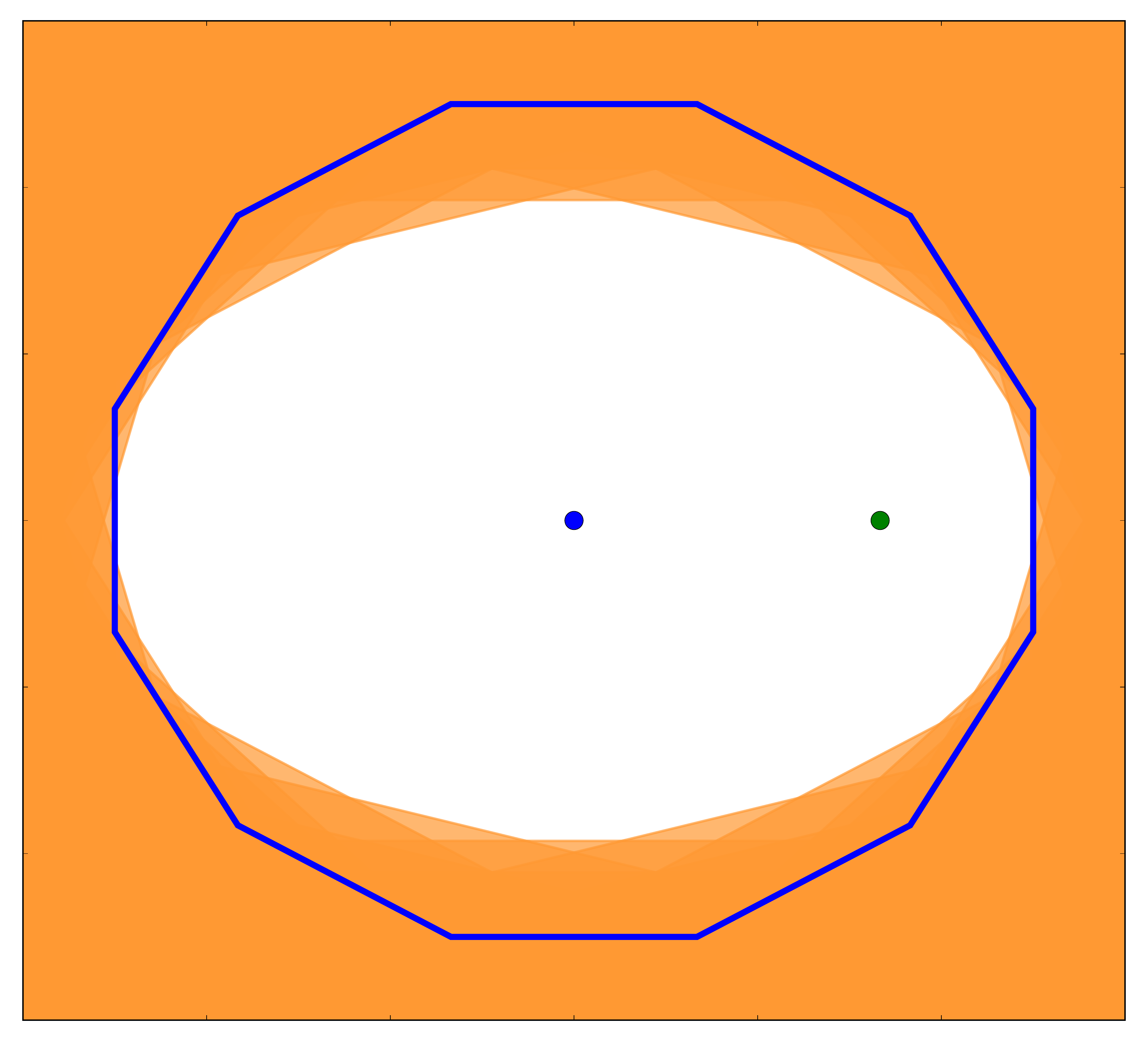}
              \hspace{1.8cm} 2k. [12, 2/3, 0]
            \end{center}
          \end{minipage} 
        \end{tabular}
        \caption{(Color online) The coexistent regions on reflecting hyperplanes in even-sided regular polygon effect spaces. The blue lines indicate the hyperplane and the blue points are at the center of the polygons. The green points are fixed effects $e$ and the white regions are coexistent region with $e$. The captions [$n, s, v$] under each figure say the number of edge, the ratio of $e$ from the origin to the edge, and that to the vertexes, respectively. The upper series (from a. to d.) show how the coexistent region changes by $e$ closing to middle of the edge. The middle series (a. and from e. to g.) show how it does by $e$ closing to the vertex. The lower series (from h. to k.) show the quantum limit, in other words how the region change by the increase of the number of edge.}
        \label{fig:CERs}
      \end{center}
    \end{figure*}

    We show the coexistent regions with the fixed effect $e$ in FIG. \ref{fig:CERs}.
    This theorem claims the closer an effect is to the boundary, the smaller its coexistent region is. It is important to see that for the extremal effect its coexistent area vanishes.
    
    Now, we consider the case $n \rightarrow \infty$. The limit allows the angles $\frac{2k_i}{n}\pi$ to be re-parametrized by arbitrary real numbers $\theta_i$. Thus, we obtain 
    \begin{eqnarray}
      (f_x-e_x)\cos\theta_1 + (f_y-e_y)\sin\theta_1 & + & \nonumber\\
      (f_x+e_x)\cos\theta_2 + (f_y+e_y)\sin\theta_2 && \leq 1 \label{eq:ce_eq_lim}.
    \end{eqnarray}
    The sum of the first two terms equals to $\|(e_x,e_y)+(f_x,f_y)\|$, and that of the last two terms equals to $\|(e_x,e_y)-(f_x,f_y)\|$. Therefore, we regain the inequality (\ref{eq:qce_mod}). This result shows that the quantum limit of Theorem \ref{th:ce} coincides with Theorem \ref{th:qce}.
    
  \subsection{The volume of coexistent region vanishes for extreme effects}
    We study a general effect space with a reflecting hyperplane in this section. The previous theorems show that for the regular polygon theories the volume of coexistent regions for the nontrivial extreme effects vanishes. We generalize this property to arbitrary effect spaces with reflecting hyperplane. An effect space we consider is embedded in a vector space which equips a natural volume measure (Lebesgue measure).
    
    Before showing the theorem, we introduce a term ``edge'' and two lemmas.
    \begin{definition}
      A line segment $[e_1,e_2]:=\{g\in E\,\, | \,\,g=pe_1+(1-p)e_2,\, p\in[0,1]\}$ is called an {\em edge} if and only if $g$ has no decomposition such that $g = r_1e_1+r_2e_2+q_1f_1+q_2f_2+\cdots+q_{d+1}f_{d+1}$ where $f_1,f_2,\cdots f_{d+1}(\not=e_1,e_2)$ is extremal, $0<r_1,r_2,q_1,q_2,\, 0\leq q_i (i\geq3),\, r_1+r_2+\sum q_i=1$, and $d$ is dimension of effect space.
    \end{definition}
    This definition claims that an edge is one-dimensional boundary of an effect space. The maximum of suffixes is obtained by Carath\'{e}odory's theorem, which says any point in $d$-dimensional convex hull lies in convex hull of $d+1$ or fewer points on original convex hull. The following lemma is a general property about edges on a general effect space.
    \begin{lemma}
      \label{lm:edge1}
      For any extremal effects $e$, if $[o,e]$ is edge then $\underline{E}(e) = [o,e]$.
      \begin{proof}
        If there exist an effect $e'$ such that $e'\in \underline{E}(e),\, e'\not\in[o,e]$ then $\bar{e}+e'\in E$ and $\bar{e}+e'\not\in[\bar{e},u]$. This is equivalent to $\overline{(\bar{e}+e')}=e+\bar{e'}\in E$, $e+\bar{e'}\not\in[o,e]$. The convex sum $\frac{1}{2}(e+\bar{e'})+\frac{1}{2}(e') = \frac{1}{2}o+\frac{1}{2}e$ is on $[o,e]$, since $[o,e]$ is edge. This contradicts against the fact that both $e+\bar{e'}, e'$ have convex decomposition with other extremal points.
      \end{proof}
    \end{lemma}
    The next lemma shows a property of edges on an effect space with a reflecting hyperplane.
    \begin{lemma} 
      \label{lm:edge2}
      For all extremal effects $e$ (except for $o, u$) on a reflecting hyperplane, the line segments $[o,e], [e,u]$ are edges.
      \begin{proof}
        We only show that $[0,e]$ is an edge. If an effect $g'$ on the line segment has the decomposition then we have an equation
        \begin{eqnarray}
          g' &:=& \frac{q_1f_1+q_2f_2+\cdots+q_{d+1}f_{d+1}}{q_1+q_2+\cdots+q_{d+1}} \nonumber\\
          &=& \frac{(p-r_1)o+(1-p-r_2)e}{1-r_1-r_2}.
        \end{eqnarray}
        This means that the effect $g'$ is on the convex hull of effects $f_1, f_2, \cdots, f_{d+1}$. As the effect space has a reflecting hyperplane, the convex hull is also on reflecting hyperplane. Therefore, $g'$ is on the hyperplane. On the other hand, $g'$ is on the edge $[o, e]$, as the r.h.s. of the equation shows. $e$ is also an effect on the hyperplane, thus we get $g' = e$. It contradicts against the extremality of $e$.
      \end{proof}
    \end{lemma}
    Now we show a desired theorem.
    \begin{theorem}
      If an effect space whose dimension is more than three has a reflecting hyperplane, then the the volume of the coexistent region with respect to any nontrivial extremal effect is vanishing.
      \begin{proof}
        Lemma \ref{lm:edge2} says that the lower set of an extremal effect is an edge $[0,e]$, and the upper set of its complement is an edge $[\bar{e},u]$. Both edges are one-dimensional region shown in the lemma \ref{lm:edge1}, therefore their Minkovski's sum are at most two-dimensional body. Thus we conclude that the coexistent volume is $0$ since the effect space is at least three-dimensional body.
      \end{proof}
    \end{theorem}
    
    Note that we restricted the dimension of effect space is not two in the last theorem. This is because the effect space whose dimension is two must be equivalent to the classical bit space. The bit system has two extremal effects, and all pairs of effects (not only extremal ones) are coexistent.
    
    We may address the fact that the converse of the theorem is not true. Let us consider an effect space corresponding to a hexagon state space, which has extremal effects $o,e^6_i,u \, (i=0,1,\cdots,5)$. We choose two of them, $e^6_0,e^6_3 = (\pm\frac{1}{2},0,\frac{1}{2})$, and displace them $\pm\delta \,(0<\delta<\frac{1}{2})$ to $z$ direction respectively. This new effect space still satisfies the effect space axioms, since $e'^6_0+e'^6_3=u$, but it has no more reflecting hyperplane. The coexistent region with respect to $e'^6_0$ is the Minkowski's sum of $\underline{E}(e'^6_0)$ and $\underline{E}(e'^6_3)$, and both of them are 1 dimensional region. Thus, we conclude that this is a counter example.

\section{Conclusion and Outlooks}
\label{sec:CandO}
  We have shown that the existence of a reflecting hyperplane enables a generalization of the unbiased effects in the quantum theory and the coexistence problem between such generalized unbiased effects are treatable. In fact, we obtained a necessary and sufficient condition for a pair of unbiased effects in the even-sided regular polygon theories to be coexistent. This result, in a certain limit, reproduces a low-dimensional version of Busch's result. We showed that the existence of a reflecting hyperplane is related to the point symmetry of the corresponding state space. In this sense, this property is rather specific. In fact, one can show that the effect spaces of the trit and qutrit do not have reflecting hyperplane. We expect that the existence of the reflecting hyperplane is related to a certain sort of minimality of the information capacity of the system.
  
\begin{acknowledgments}
  The authors would like to thank Ikko Hamamura for his comments on an earlier version of this paper and his other proof of proposition \ref{prop:ESandSS}. We also thank Yui Kuramochi for pointing out that the generalization have some difficulties. Finally, we are grateful to Takayuki Miyadera for his direction of this paper as our supervisor.
\end{acknowledgments}

\bibliographystyle{apsrev4-1}
\bibliography{GeneralizedProbabilityTheory}

\end{document}